\documentclass[prb]{revtex4}

\usepackage{graphicx}%
\usepackage{dcolumn}
\usepackage{amsmath}

\makeatletter
\def\btt#1{\texttt{\@backslashchar#1}}%
\DeclareRobustCommand\bblash{\btt{\@backslashchar}}%
\makeatother

\begin{document}

\preprint{HEP/123-qed}

\title[Short Title]{
Superconducting anisotropy and\\
evidence for intrinsic pinning in single crystalline MgB$_2$
}

\author{Ken'ichi Takahashi, Toshiyuki Atsumi, and Nariaki Yamamoto}

\affiliation{%
Department of Physics and Electronics, 
Osaka Prefecture University, 
Sakai, Osaka 599-8531, Japan
}%

\author{Mingxiang Xu and Hideaki Kitazawa}
\affiliation{
Nanomaterials Laboratory, 
National Institute for Materials Science (NIMS), 
1-2-1 Sengen, Tsukuba,                  
Ibaraki 305-0047, Japan 
}%

\author{Takekazu Ishida}
 \email{ishida@center.osakafu-u.ac.jp}
\affiliation{
Department of Physics and Electronics, 
Osaka Prefecture University, 
Sakai, Osaka 599-8531, Japan}

\date{February 28, 2002}

\begin{abstract}
We examine the superconducting anisotropy $\gamma_c = (m_c / m_{ab})^{1/2}$ of a metallic high-$T_c$ superconductor MgB$_2$ by measuring the magnetic torque of a single crystal. The anisotropy $\gamma_c$ does not depend sensitively on the applied magnetic field at 10 K. We obtain the anisotropy parameter $\gamma_c = 4.31 \pm 0.14$. The torque curve shows the sharp hysteresis peak when the field is applied parallel to the boron layers. This comes from the intrinsic pinning and is experimental evidence for the occurrence of superconductivity in the boron layers. 
\verb
\pacs{74.25.Ha,74.60.-w}

\end{abstract}

\pacs{ 74.25.Ha,  74.60.-w}
\maketitle


%
\section{Introduction}               

Since the discovery of superconductivity in MgB$_2$ \cite{Nagamatsu} considerable progress has been made in determining the physical properties of this material.
The new materials are metallic and hence promising in considering the applications in the various fields. 
The extreme high-$T_c$ (39 K) gives a doubt whether or not the superconductivity can be explained within a conventional BCS framework. 
Tunneling studies show that the material is reasonably isotropic and has a well developed $s$-wave energy gap. \cite{Schmidt,Sharoni}
The anisotropy parameter $\gamma_c=(m_c/m_{ab})^{1/2}$ of MgB$_2$ in the literature ranges from 1.2 to 9 in polycrystalline samples. \cite{Buzea}
Therefore, it is highly desirable to investigate the fundamental properties of MgB$_2$ by using a single crystal. 
Recently, Xu et al. \cite{Xu} succeeded in synthesizing the single crystals by the vapor transport method and reported the superconducting properties of MgB$_2$. 

The first principle calculation by Kortus et al. \cite{Kortus}, by Choi et al. \cite{Choi}, and by Yildirim et al. \cite{Yildirim} suggested that the boron layers govern the superconductivity.
The boron isotope effect on $T_c$ supports this idea.\cite{Bud'ko}
However, the experimental confirmation of the superconductivity in the boron layers is not thoroughly convincing so far. 

Torque is a sensitive tool for probing the various kinds of anisotropy, and has been successfully applied to investigate the highly anisotropic high-$T_c$ cuprates.  
The electronic anisotropy of high-$T_c$ cuprates was investigated by Farrell {\it et al.} \cite{Farrell} and Ishida {\it et al.}\cite{Ishida97B} 
The high-$T_c$ cuprates are characterized by the extreme electronic anisotropy $\gamma_c=\sqrt{m_c/m_{ab}} = 7 - 200$ as well as the layered structure of superconductivity. 
The superconductivity is governed by the CuO$_2$ layers or by the CuO$_2$ bilayers. 
An alternative stacking of the CuO$_2$ layer and the blocking layer is a key concept both in crystal structure and in the occurrence of the superconductivity. 
This is also the origin of the intrinsic pinning for vortices in high-$T_c$ superconductors. 
The MgB$_2$ structure consists of an alternative stacking of the boron layer and the magnesium layer, too.
It is of special interest to investigate the similarity and dissimilarity between this new material MgB$_2$ and the cuprate superconductors.

It is crucial to discriminate the intrinsic pinning from various other pinning sources.
The magnetic torque has an advantage to see the vortex pinning as a function of angle with respect to a crystalline axis.
Since the intrinsic pinning works effectively when the field is almost in parallel to the CuO$_2$ planes, it manifests as a hysteresis peak in torque.\cite{Tachiki,Ishida01}
Therefore, a torque is a sensitive probe for sensing an intrinsic (directional) pinning.

In this paper, we report the electronic anisotropy of MgB$_2$ by means of the magnetic torque.
We also describe evidence for the intrinsic pinning in MgB$_2$ when the field is applied 
parallel to the boron layers. This ensures that the superconductivity occurs in the boron layers.

\section{Experimental}

\subsection{Sample}

The details of the single crystal preparation of MgB$_2$ is reported by Xu et al.\cite{Xu} 
They reported that the onset temperature of superconductivity is 38.6 K.
Because of the severe volatility of Mg and the high melting point of B, MgB$_2$ single crystals were grown in a closed system. The starting material of Mg is 99.99\% and B is 99.9\% in purity. 
A molar ratio of 1:1.9 of the starting materials was sealed inside a molybdenum crucible of internal diameter 10 mm, length 60 mm by the electron beam welding. 
The molybdenum crucible was used in a high frequency induction furnace. 
The crucible was first heated to 1400 $^\circ$C at a rate of 200 $^\circ$C/h and kept for 2 h, then slowly cooled to 1000 $^\circ$C at a rate of 5 $^\circ$C/h, and finally to room temperature by switching off the power. 
A plate-like single crystal of MgB$_2$ ($\sim$ 0.11 mm$^2$ $\times$ 14 $\mu$m) is used for the torque measurements.
The $c$-axis is perpendicular to the plate. 
The sample weight is too tiny to measure by our electronic balance of resolution 0.01 mg.

\subsection{Torque}

A split-type superconducting magnet changes a magnetic field continuously from $-60$ kG to +60 kG, and has a variable temperature insert from 4 K to 300 K. A torque detection system is attached on the top of the insert. 
A phosphor bronze string hangs a balance arm, a feedback coil, and a quartz sample rod in a helium-gas atmosphere from the main flange. The feedback coil (300 turns of 0.075-mm Cu wire) is located in a special hollow cylindrical NdFeB permanent magnet (Magnetic Solutions) of a transverse field of 5000 G to 8000 G. A sample torque can be cancelled by a controlled torque given by a torque detection mechanism at the top of the insert. We use an optical position sensor to maintain the sample direction. The sample can be rotated by a stepper motor (a resolution of $0.0036^\circ$). 
The torque dynamic range is from $-10^{3}$ dyncm to $+10^3$ dyncm with a sensitivity of $10^{-3}$ dyncm.

\section{Results and discussions}

We measured the torque of a single crystalline MgB$_2$ as a function of $\theta_{c}$. 
The angular step was chosen as 0.5 degrees. 
The irreversible torque was extracted by $\tau_{irr}(\theta)=(\tau_{dec}(\theta)-\tau_{inc}(\theta))/2$ and the reversible torque was obtained as $\tau_{rev}(\theta)=(\tau_{inc}(\theta)+\tau_{dec}(\theta))/2$ where $\tau_{inc}(\theta)$ and $\tau_{dec}(\theta)$ are the torques as a function of increasing and decreasing angle, respectively.

\begin{figure}[btp]
\begin{center}
\leavevmode
\includegraphics[width=0.8\linewidth]{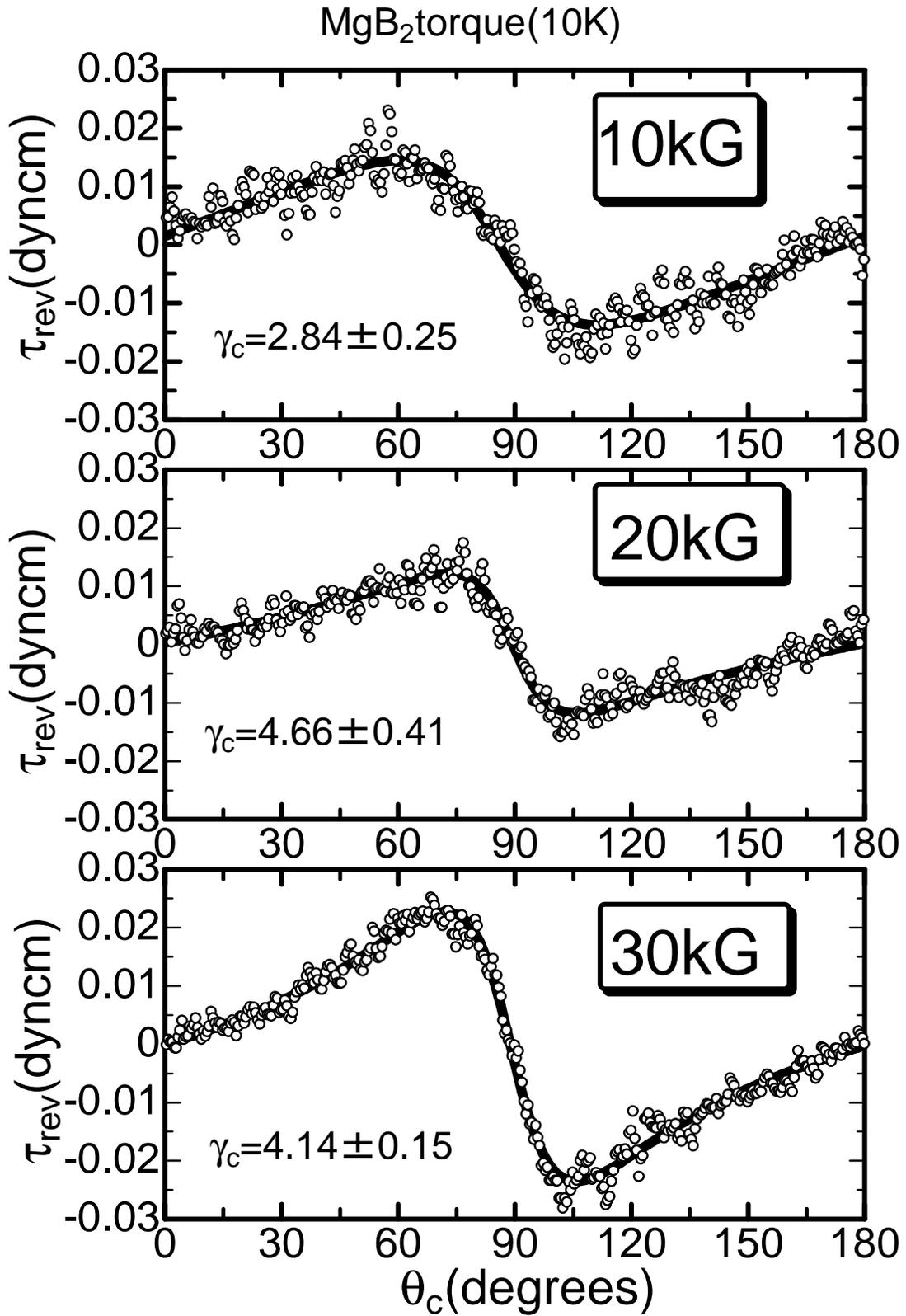}
\vspace*{1cm}
\caption{
The reversible torque $\tau_{rev}$ of MgB$_2$ as a function of angle $\theta_{c}$ at 10 K (10 kG, 20 kG, 30 kG). 
The reversible torque is determined as $\tau_{rev}(\theta)=(\tau_{inc}(\theta)+\tau_{dec}(\theta))/2$ where $\tau_{inc}(\theta)$ and $\tau_{dec}(\theta)$ are the torques as a function of increasing and decreasing angle. The torque curves are analyzed the Kogan model by fixing $\eta H_{c2}=60$ kG (see $\gamma_c$ of the figures).
}
\label{fig:01}
\end{center}
\end{figure}

\begin{figure}[btp]
\begin{center}
\leavevmode
\includegraphics[width=0.8\linewidth]{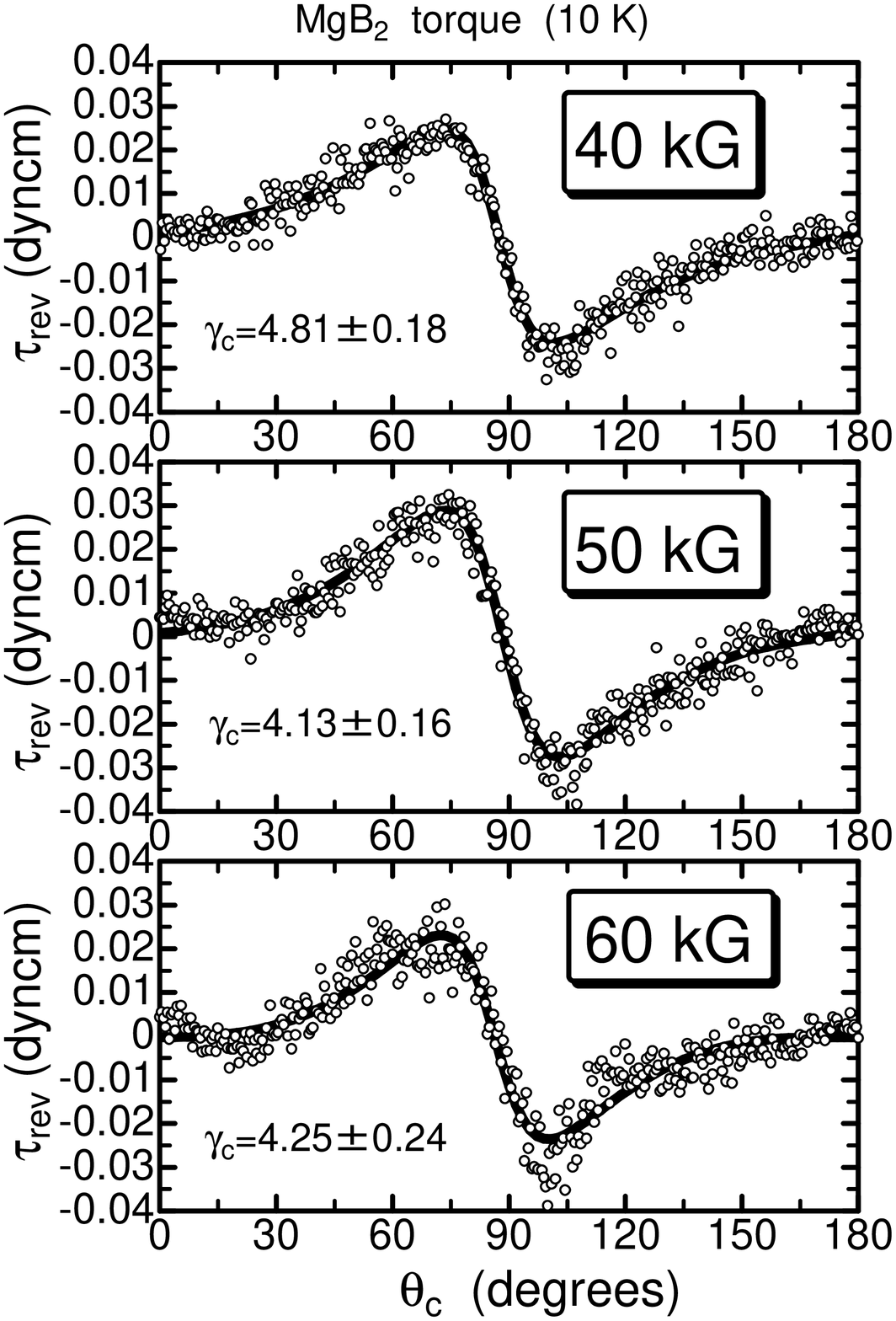}
\vspace*{1cm}
\caption{
The reversible torque $\tau_{rev}$ of MgB$_2$ as a function of angle $\theta_{c}$ at 10 K (40 kG, 50 kG, 60 kG). 
The reversible torque is determined as $\tau_{rev}(\theta)=(\tau_{inc}(\theta)+\tau_{dec}(\theta))/2$ where $\tau_{inc}(\theta)$ and $\tau_{dec}(\theta)$ are the torques as a function of increasing and decreasing angle. The torque curves are analyzed the Kogan model by fixing $\eta H_{c2}=60$ kG (see $\gamma_c$ of the figures).
}
\label{fig:02}
\end{center}
\end{figure}

In Figs.~\ref{fig:01} and \ref{fig:02}, we show the reversible torque $\tau_{rev}$ of MgB$_2$ at 10 K (10 kG, 20 kG, 30 kG) and the reversible torque $\tau_{rev}$ of MgB$_2$ at 10 K (40 kG, 50 kG, 
60 kG), respectively. 
The shape of the torque curves in Figs.~\ref{fig:01} and \ref{fig:02} is similar to the torque curve of YBa$_2$Cu$_3$O$_{7-\delta}$.\cite{Ishida97B}
This indicates that the MgB$_2$ superconductor is anisotropic with respect to the $c$ axis.
In the three-dimensional anisotropic London model in the mixed state, the angular dependence of the torque is given by Kogan \cite{Kogan} as 
\begin{eqnarray}
\tau_{rev}(\theta_{c})=
{{\phi_0 H V}\over {16\pi\lambda^2}}
{{\gamma_{c}^2-1}\over \gamma_{c}^{1/3}}
{{\sin 2\theta_{c}}\over {\epsilon(\theta_{c})}} 
\ln \{{{\gamma_{c} \eta H_{\rm c2}^{\bot}} 
\over {H\epsilon(\theta_{c}})}\} 
\label{eq:eq01}
\end{eqnarray}
where $\epsilon(\theta_{c}) = (\sin^2\theta_{c}+\gamma_{c}^2 \cos^2\theta_{c})^{1/2}$, $\theta_{c}$ is the angle between the applied field and the $c$ axis, $\gamma_{c}=\sqrt{m_c/m_{ab}}$, $H_{\rm c2}^{\bot}$ is the upper critical field perpendicular to the $ab$ plane ($\eta \sim 1$), $V$ is the sample volume. 
This equation is relatively simple, and has frequently been employed in the literature to analyze the electronic anisotropy of various high-$T_c$ cuprates.\cite{Farrell,Ishida97B,Ishida98} 
The computer fitting of $\tau_{rev}$ to the Kogan model gives the anisotropy parameter $\gamma_{c}=2.8-4.8$ where $\eta H_{c2}$ is fixed to 60 kG.\cite{Xu} 

\begin{figure}[btp]
\begin{center}
\leavevmode
\includegraphics[width=0.9\linewidth]{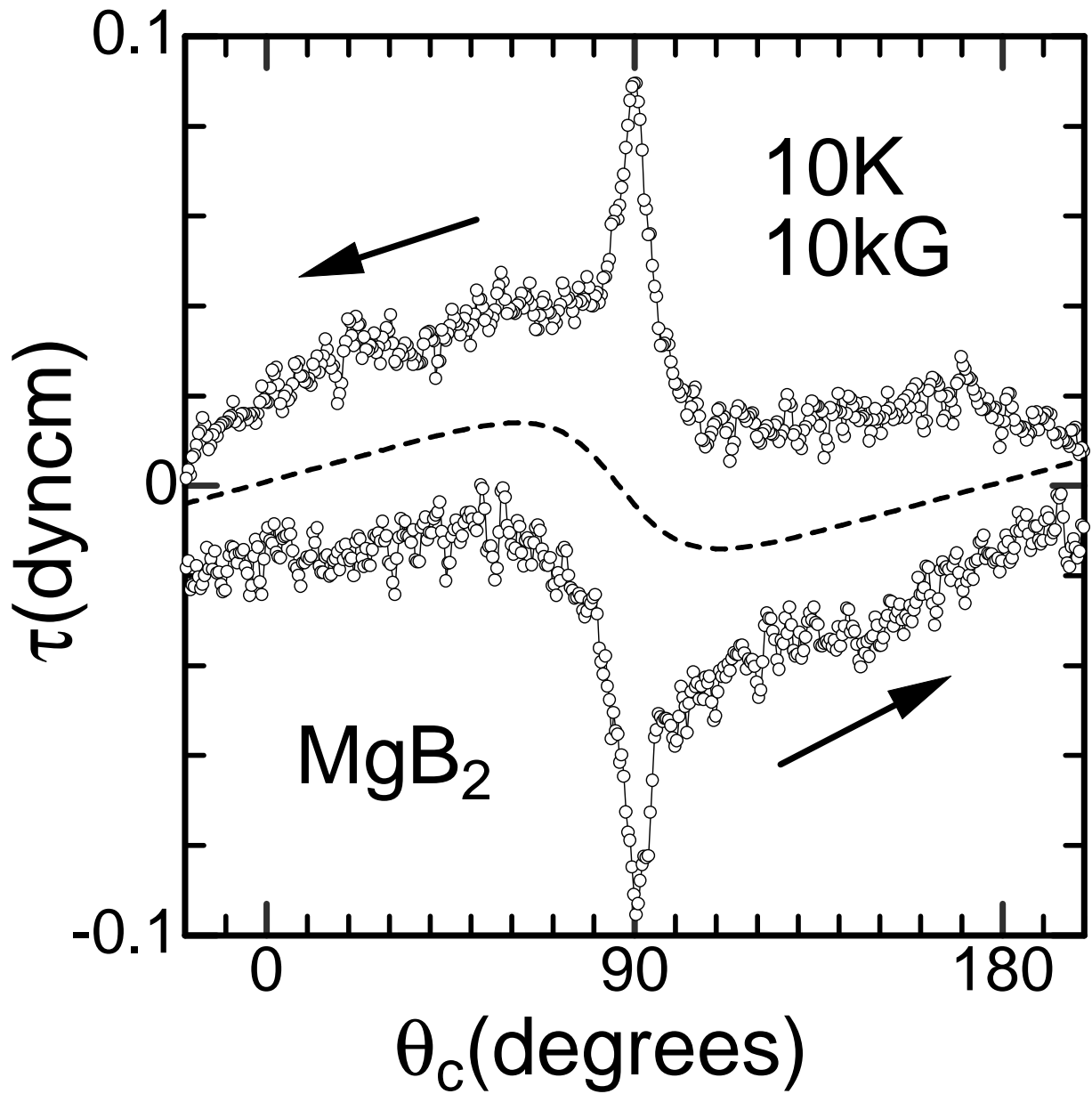}
\vspace*{1cm}
\caption{
The torques as a function of increasing as well as decreasing angle are presented in 10 kG at 10 K.
The dashed line is the reversible component (see top figure of Fig.~\ref{fig:01}). 
The peak appeared near $90^\circ$ represents the manifestation of intrinsic pinning 
in this novel superconductor.
This is clear evidence for layered superconductivity in the boron layers. 
}
\label{fig:03}
\end{center}
\end{figure}

As shown in Fig.~\ref{fig:03}, the torque curve measured in 10 kG at 10 K has a hysteresis against the angle scans. 
The torque as a function of increasing as well as decreasing angle has a sharp peak at $\theta_c \simeq 90^\circ$. 
This is well known as an intrinsic pinning peak for the high-$T_c$ cuprates.\cite{Tachiki,Ishida97B}
The dashed line is the reversible component (see the fitted line in the top figure of Fig.~\ref{fig:01}). 
The peak appeared near $90^\circ$ represents the manifestation of intrinsic pinning in this novel superconductor. 
This is experimental confirmation of the superconductivity in the boron layers.\cite{Kortus,Choi,Yildirim} 
The remarkable hysteresis of the torque curve give rise to the uncertainty in the reversible torque obtained by  $\tau_{rev}(\theta)=(\tau_{inc}(\theta)+\tau_{dec}(\theta))/2$. 
Actually, $\gamma_c$ in 10 kG is appreciably less than those in other fields at 10 K. 
A tiny intrinsic pinning peak can be seen in the torque curve at 20 kG. 
Note that the torque curves are almost reversible for 30 kG, 40 kG, 50 kG, and 60 kG.

The $\gamma_c$ is almost independent of field between 20 kG and 60 kG at 10 K.
We omitted $\gamma_c$ in 10 kG to obtained the averaged $<\gamma_c > = 4.31 \pm -0.14$.
This is somewhat larger than those reported in the literature.\cite{Buzea}

Angst et al. \cite{Angst} also reported the torque measurement of MgB$_2$.
Their measurement regime is complimentary to us. 
They mainly use the torque as a sensitive means of determining $H_{c2}$ 
as a function of angle $\theta_c$.
However, the determination of $H_{c2}$ is dependent on the criterion of the torque onset, 
and their criterion contains a target parameter $\gamma_c$. 
The upper critical field $H_{c2}(\theta_c)$ thus obtained is analyzed by the effective mass model.
They obtained $\gamma=6$ at 15 K and $\gamma=2.8$ at 35 K. 
They also used the Kogan model to analyze the torque data and found the field dependence of $\gamma_c$.
They interpreted it in terms of the double gap structure of MgB$_2$. 
We note that this is not the case in our torque measurements because $\gamma_c$ is almost independent of field at 10 K.

We attempted to measure the temperature dependence of the torque curve in 60 kG, but the signal noise ratio is not satisfactory enough to analyze the torque data.  
This is due to the smallness of the MgB$_2$ crystal.
It is inconclusive on the temperature dependence of $\gamma_c$. 
Measurements using a larger crystal are desirable for the detailed studies of the MgB$_2$ system.

\begin{figure}[btp]
\begin{center}
\leavevmode
\includegraphics[width=0.9\linewidth]{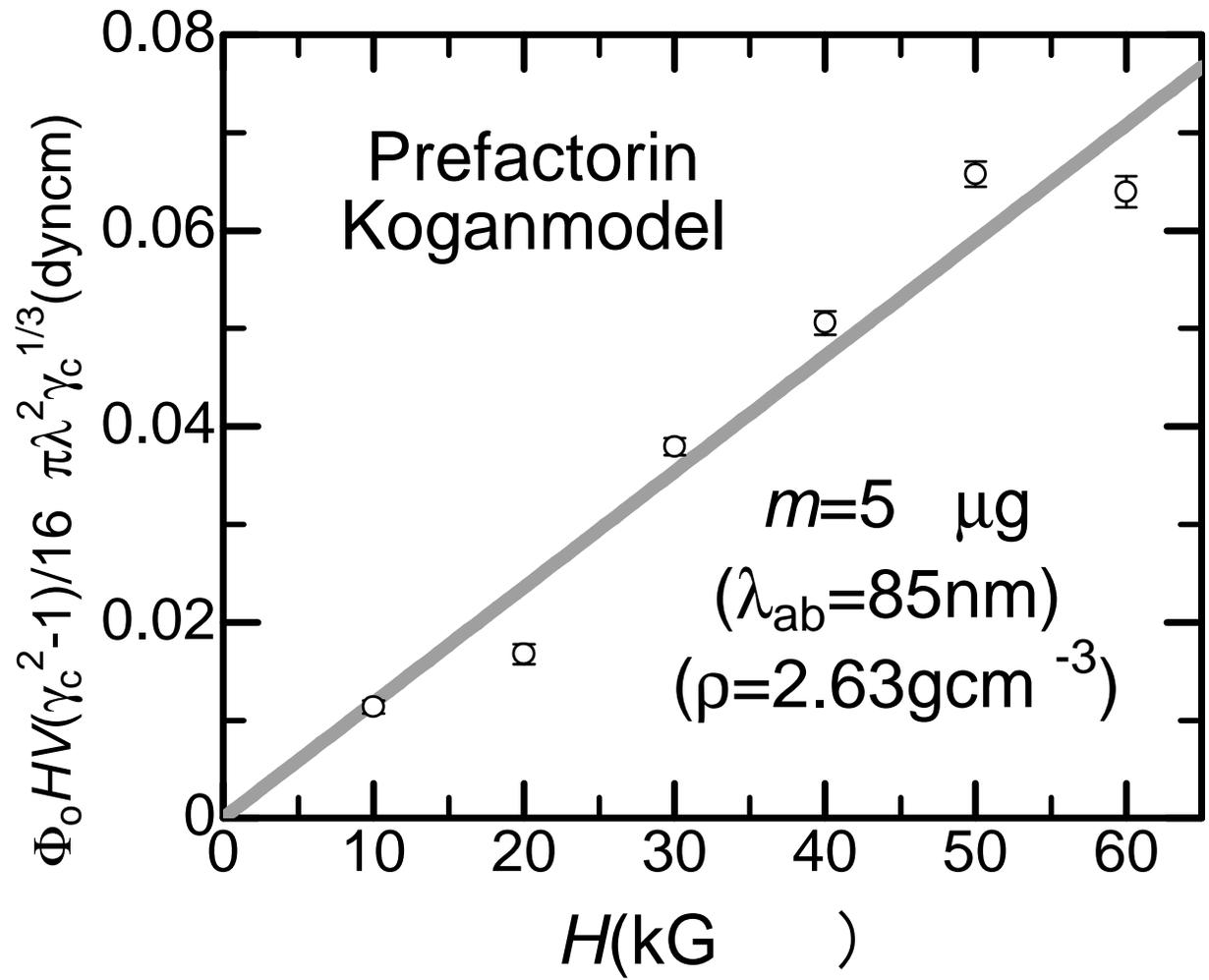}
\vspace*{1cm}
\caption{
A prefactor ${{\phi_0 H V}({\gamma_{c}^2-1)/ {16\pi\lambda^2}}\gamma_{c}^{1/3}}$ of the Kogan formula as a function of $H$. 
From the slope, the sample mass is estimated as approximately 5 $\mu$g.
}
\label{fig:05}
\end{center}
\end{figure}

Finally, we estimate a weight of our MgB$_2$ sample from the torque amplitude.
In the Kogan model, we obtain a prefactor of ${{\phi_0 H V}/ {16\pi\lambda^2}}$ by the least squares fitting where the penetration depth should read as $\lambda=(\lambda_c \lambda_{ab}^2)^{1/3}$.
In Fig.~\ref{fig:05}, we plot ${{\phi_0 H V}/ {16\pi\lambda^2}}$ as a function of $H$. 
The data points are approximately fitted by a straight line.
From the slope, one determine the sample volume $V$. 
By assuming $\lambda_{ab}=85$ nm,\cite{Buzea} the density $\rho=2.63$ g/cm$^3$, and $\gamma_c=4.31$ we estimate the mass $m$ of our sample is $\sim 5 \mu$g.
We note that the mass calculated from the measured size of crystal and the density of MgB$_2$ is approximately 3.3 $\mu$g.

\section{Conclusions}

The torque curves at 10 K are almost reversible in fields larger than 20 kG. 
The electronic anisotropy of MgB$_2$ is determined by the torque as $\gamma_c=4.31 \pm 0.14$.
The new superconductor exhibits an intrinsic pinning at 10 K in 10 kG when the field is applied parallel to the boron layers. 
There is a modulation of the order parameter in MgB$_2$ along the $c$ axis. 
This indicates that the boron layers, as suggested from the first principle calculations,\cite {Choi} govern the superconductivity.
The MgB$_2$ superconductor is rather similar to the high-$T_c$ cuprates while the anisotropy of MgB$_2$ is moderate compared to the cuprates.

\begin{acknowledgments}
This work was partially supported by a Grant-in-Aid for Scientific Research (Project 12554012, Project 12874042) granted by the Ministry of Education, Science, and Culture of Japan. 
\end{acknowledgments}

\end{document}